\documentclass[a4paper]{revtex4}
\usepackage{amssymb}
\usepackage{amsfonts}
\usepackage{graphicx}
\usepackage{subfigure} %para subfiguras
\usepackage{comment} %Para poner comentarios
\usepackage{hyperref} %para poner automaticamente links a bibliografia y links a paginas web con \url{http://www.}

\newcommand{\qed}{\hfill \mbox{\raggedright \rule{.07in}{.1in}}}
\newtheorem{defi}{DEFINITION}[section]
\newtheorem{prop}{PROPOSITION}[section]
\newtheorem{teo}{THEOREM}[section]
\newtheorem{prob}{PROBLEM}[section]
\newtheorem{corol}{COROLLARY}[section]

\begin{document}
\title{Quantum tomography meets dynamical systems and bifurcations theory}
\author{D. Goyeneche}
\email{dardo.goyeneche@cefop.udec.cl}
\affiliation{Departamento de Fis\'{i}ca, Universidad de Concepci\'{o}n, Casilla 160-C, Concepci\'{o}n, Chile\\Center for Optics and Photonics, Universidad de Concepci\'{o}n, Casilla 4016, Concepci\'{o}n, Chile}
\author{A. C. de la Torre}
\affiliation{Departamento de F\'{i}sica, Universidad Nacional de Mar del Plata\\Dean Funes 3350, 7600 Mar del Plata, Argentina IFIMAR-CONICET}

\begin{abstract}
A powerful tool for studying geometrical problems in Hilbert space
is developed.  In particular, we study the quantum pure state
tomography problem in finite dimensions from the point of view of
dynamical systems and bifurcations theory. First, we introduce a
generalization of the Hellinger metric for probability distributions
which allows us to find a geometrical interpretation of the quantum
state tomography problem. Thereafter, we prove that every solution
to the state tomography problem is an attractive fixed point of the
so--called \emph{physical imposition operator}. Additionally, we
demonstrate that multiple states corresponding to the same
experimental data are associated to bifurcations of this operator.
Such a kind of bifurcations only occurs when informationally
incomplete set of observables are considered. Finally, we prove that
the physical imposition operator has a non--contractive Lipschitz
constant 2 for the Bures metric. This value of the Lipschitz
constant manifests the existence of the quantum tomography problem for pure
states.\vspace{0.3cm}

\end{abstract}
\maketitle
\date{\today}
\section{Introduction}
In earlier times of quantum mechanics W. Pauli posed an intriguing
comment in a footnote \cite{Pauli}. He mentioned that the problem to
univocally determine the wave function of a particle from the
knowledge of position and momentum distributions was not fully
explored. This question, known as \emph{Pauli problem}, is the
origin of the quantum state reconstruction problem currently
denominated \emph{quantum state tomography}. The importance of
quantum tomography increased in the last years due to its important
role in the so called quantum technologies (quantum teleportation,
quantum computing, quantum metrology, etc). Concerning finite
dimensional Hilbert spaces some progresses have been found. For
example, the statistics obtained from
measuring three probability distributions associated to the spin
observables $S_x, S_y$ and $S_z$ is enough to reconstruct almost all
quantum states, up to a null measure set in Hilbert space
\cite{Weigert2}. Furthermore, Moroz and Perelomov have proven that the
statistics collected from any three observables is not enough to
reconstruct any quantum state in dimensions $d\geq9$ \cite{Moroz}.
More recently, from considering compressed sensing it has been proven
that rank $r$ observables can be reconstructed from $O(rd\log^2d)$
measurements \cite{Gross}. Also, an informationally complete set of
observables for pure states has been found, where the number of
rank--one projective measurements is linear with the dimension
\cite{Goyeneche4}. On the other hand, the Pauli problem also exists in infinite
dimensional Hilbert spaces \cite{Corbett,Stulpe,Krahmer,Corbett2} where it is close
related to radar signals theory \cite{Jaming,Orlowski}.

In this work, we prove that dynamical systems and bifurcations
theory are suitable tools to have a deep understanding of several
geometrical problems existing in Hilbert spaces such as quantum
state reconstruction problem, mutually unbiased bases, symmetric
informationally complete POVM and real and complex Hadamard
matrices. Here, our explanations and proves are mainly focused on
the quantum state reconstruction problem. Despite this, the
extension to any of the problems mentioned above is straightforward.
In Section II we present an introduction to the quantum state
reconstruction problem. The aim of this section is to present the
minimum of contents required to understand this work. Expert readers
may skip this section. In Section III we define some metrics which
are required to prove the convergence of the sequences involved in
our method. Here, we introduce the novel concept of
\emph{distributional metric} which is a generalization of the well
known Hellinger metric. This new metric allows us to compare the distance between probability distributions contained in quantum states. Additionally, from a simple geometrical argument involving this metric we intuitively explain why the Pauli problem exists. In
Section IV we define the physical imposition operator and we prove
that every solution to the quantum state reconstruction problem is
an attractive fixed point of this operator. In Section V we explain
the connection between quantum state tomography and bifurcation
theory. Finally, in Section VI we resume our results and conclude.
The most important proofs of this work are given in Appendix I.

\section{An introduction to the quantum state reconstruction problem}\label{QM}
As a fundamental principle, in quantum mechanics we cannot
reconstruct a state from a single observation. This is a direct
consequence of the collapse of the wave function, an irreversible
process that destroys the information contained in the original
system. Another way to explain this fact is provided by the
\emph{non--cloning theorem} \cite{Wootters2}, which states that a
single quantum system cannot be perfectly copied. Therefore, in
order to reconstruct a state we require to have an ensemble of
identical quantum systems. Additionally, we are able to make a
single measurement in each particle of the ensemble. In this work we restrict our attention to pure states defined in finite dimensional Hilbert spaces and we consider traditional Von
Neumann observables. Generalized POVM measurements will be sporadically mentioned but not considered in our method. Consequently, the information gained in the laboratory will be
provided by rank--one projective measurements that can be sorted in
orthogonal bases. We use standard Hilbert space notation instead of the Dirac one frequently used in quantum mechanics.

Let us assume a $d$ dimensional Hilbert space $\mathcal{H}$, an
orthonormal basis $\{\varphi_k\}\in\mathcal{H}$ and a physical
system prepared in the state $\Phi$. The Born's rule tells us that
\begin{equation}\label{Born}
p_k=|\langle\varphi_k,\Phi\rangle|^2,
\end{equation}
where ${k=0,\dots,d-1}$ is the normalized probability distribution
of the eigenvalues of the observable associated to the basis. This
equation allows us to make two important processes:
\begin{enumerate}
  \item [1)] Decode the information stored in an unknown quantum state $\Phi$.
  \item [2)] Reconstruct the quantum state $\Phi$ from a set of (known) probability distributions.
\end{enumerate}
The problem of choosing a \emph{suitable} set of orthonormal bases in order to solve $2)$ represents an important open problem in quantum mechanics. Here, the concept of informationally complete set of observables naturally arises.
\begin{defi}[Informationally completeness]
A set of $m$ observables $A^1,\dots,A^m$ having eigenvectors bases $\{\varphi^j_k\}$ is informationally complete if every quantum state $\Phi$ has associated a different set of probability distributions
\begin{equation}\label{Born1}
p^j_k=|\langle\varphi^j_k,\Phi\rangle|^2,
\end{equation}
where $k=0,\dots,d-1$ and $j=1,\dots,m$.
\end{defi}
From Eq.(\ref{Born1}) we realize why the determination of a quantum
state $\Phi$ from observable quantities $\{p^j_k\}$ is so difficult:
the set of equation is non--linear. In order to minimize resources
in the reconstruction process it is very important to solve the
following problem:

\begin{prob}\label{problem}
Which are the optimal observables required to univocally determine any pure state $\Phi$?
\end{prob}
We say optimal in the sense of minimizing the following three quantities:
\begin{enumerate}
  \item The number of observables.
  \item The redundancy of information.
  \item The propagation of errors.
\end{enumerate}
In the case of mixed states it is very clear that mutually unbiased
bases \cite{Ivanovic,Wootters} are optimal sets of observables in
the above way (at least in prime power dimensions). However, for
pure states it is not clear what geometrical properties should be
satisfyed. A few partial results are known in the literature
regarding these sets. For example, it has been proven that any set
of three orthogonal bases is not enough to solve Problem
\ref{problem} in any dimension $d\geq9$ \cite{Moroz}. This means
that four bases determine a \emph{weakly} informationally complete set of
measurements \cite{Flammia}. Also, preliminary studies indicate that
a fixed set of four bases is enough to reconstruct any $d$
dimensional pure state up to a null measure set of dimension $d-2$
and that five adaptative bases are enough for any pure state in
every dimension $d$ \cite{Goyeneche4}. The general solution of
Problem \ref{problem} given a fixed number of bases is still open.

 Interestingly, a lower bound $\mathfrak{m}$ has been found for
the minimum number of rank--one projectors required to form an
informationally complete set of POVM \cite{Heinosaari}:

\begin{equation}\label{lowerbound}
\mathfrak{m} > \left\{
\begin{array}{c l}
 4d-2\alpha-4 & \forall d>1\\
  4d-2\alpha-3& d\mbox{  odd, and  }\alpha=2\mbox{ mod }4\\
 4d-2\alpha-2 & d\mbox{  odd, and }\alpha=3\mbox{ mod }4
\end{array}
\right.
\end{equation}
where $\alpha$ denotes the number of ones in the binary expansion of $d-1$. In the above bounds it is assumed a POVM, what contains Von Neumann measurements as a particular case. Thus, these bounds are also valid for orthonormal bases. Indeed, these bounds generalize the results found by Moroz \cite{Moroz}. The main objective of this work is to prove that any solution to the quantum state reconstruction problem is an attractive fixed point of a non--linear operator. This concept requires to define metrics for quantum states and probability distributions. By this reason, the next section is fully dedicated to study metrics.

\section{Metrics}
In this section we define some useful metrics in quantum mechanics. We first introduce the \emph{Hellinger metric} \cite{Hellinger}, which quantify the distance between probability distributions.
\begin{defi}[Hellinger metric]
Let $A$ be an observable defined on a $d$-dimensional Hilbert space $\mathcal{H}$, $\{\varphi_k\}$ its eigenvectors basis and $\Phi,\Psi\in\mathcal{H}$. Then,
the Hellinger metric is given by
\begin{equation}\label{DistHellinger}
D_A(\Phi,\Psi)=\sqrt{\sum_{k=0}^{d-1}(\sqrt{p_k}-\sqrt{q_k})^2},
\end{equation}
where
\begin{equation}
p_k=|\langle\varphi_k,\Phi\rangle|^{2}\mbox{ and }\, q_k=|\langle\varphi_k,\Psi\rangle|^{2},
\end{equation}
for every $k=0,\dots,d-1$.
\end{defi}
It is important to remark that $D_A(\Phi,\Psi)$ does \emph{not} represent a metric for quantum states. The advantage of this notation will be appreciated in the next definition. The Hellinger metric can be also expressed as
\begin{equation}\label{DistHellinger2}
D_A(\Phi,\Psi)=\sqrt{2-2\sum_{k=0}^{d-1}\sqrt{p_k}\sqrt{q_k}}\,.
\end{equation}

The convergence criteria for the sequences of quantum states to be
studied considers the distance between sets of probability
distributions of several observables and Hellinger metric involves
only one of them. Therefore let us introduce the notion of
distributional metric as follows.
\begin{defi}[Distributional metric]
Let $A^1,\dots,A^m$ be a set of observables and
$\Phi,\Psi\in\mathcal{H}$. Then, the distributional metric is given
by
\begin{equation}\label{distvariasdistri}
\mathcal{D}_{A^1\cdots A^m}(\Phi,\Psi)=\sqrt{\frac{1}{m}\sum_{j=1}^m
D^2_{A^j}(\Phi,\Psi)}.
\end{equation}
\end{defi}
It is easy to prove that this metric is well defined. That is, the
following properties are satisfied:
\begin{enumerate}
  \item $\mathcal{D}_{A^1\cdots A^m}(\Phi,\Psi)=0$ iff $|\langle\varphi^j_k,\Phi\rangle|=|\langle\varphi^j_k,\Psi\rangle|$, for every $j=1,\dots,m$, $k=0,\dots,d-1$.
  \item $\mathcal{D}_{A^1\cdots A^m}(\Phi,\Psi)=\mathcal{D}_{A^1\cdots A^m}(\Psi,\Phi)\hspace{0.2cm}\forall\,\Phi,\Psi\in\mathcal{H}$.
  \item $\mathcal{D}_{A^1\cdots A^m}(\Phi,\Psi)\leq\mathcal{D}_{A^1\cdots A^m}(\Phi,\Xi)+\mathcal{D}_{A^1\cdots A^m}(\Xi,\Psi)\hspace{0.2cm}\forall\,\Phi,\Psi,\Xi\in\mathcal{H}$.
\end{enumerate}
Moreover, this metric is proportional to the standard metric for
real vectors in $\mathbb{R}^{md}$. If the observables
$A^1,\dots,A^m$ are non--degenerate and commute then the
distributional metric is reduced to the Hellinger metric. We are
also interested to define metrics for quantum states. That is, a
metric for complex \emph{rays} in Hilbert space, where a ray is
defined by the set $\{e^{i\alpha}\Phi\}_{\alpha\in[0,2\pi)}$ for any
$\Phi\in\mathcal{H}$. Thus, the space of complex rays is isomorphic
to the complex projective space $\mathbf{CP}^{d-1}$. Let us now
introduce a metric for quantum states.
\begin{defi}[Bures metric]
Let $\Phi,\Psi\in\mathcal{H}$. The Bures metric is given by
\begin{equation}\label{Bures}
    d(\Phi,\Psi)=\sqrt{2-2|\langle\Phi,\Psi\rangle|}.
\end{equation}
\end{defi}
The expression of the standard metric for vectors in Hilbert space
is very similar to Eq.(\ref{Bures}) but considering the real part of
$\langle\Phi,\Psi\rangle$ instead of its absolute value. The
distributional and Bures metrics have a very different meaning. However,
we can relate them by the following proposition:
\begin{prop}\label{distribucional}
Let $A^1,\dots,A^m$ be a set of $m$-observables and $\Phi,\Psi\in\mathcal{H}$. Then, the Bures metric is an upper bound for the distributional metric. That is,
\begin{equation}\label{upper1}
    d(\Phi,\Psi)\geq D_{A^1\dots A^m}(\Phi,\Psi),\,
     \forall\,\Phi,\Psi\in\mathcal{H}.
\end{equation}
Proof:
\end{prop}
Let $A^1,\dots,A^m$ be a set of observables and $\{\varphi^j_k\}$
its eigenvectors base, where $k=0,\dots,d-1$. Let
$\Phi,\Psi\in\mathcal{H}$ and consider the expansions
\begin{equation}
  \Phi=\sum_{k=0}^{d-1} \sqrt{p^j_k}e^{i\alpha^j_k}\varphi^j_k\hspace{0.3cm}\mbox{ and }\hspace{0.3cm}
  \Psi=\sum_{k=0}^{d-1} \sqrt{q^j_k}e^{i\beta^j_k}\varphi^j_k.
\end{equation}
Using the triangular inequality we find that
\begin{equation}
  |\langle\Phi,\Psi\rangle|=\left|\sum_{k=0}^{d-1}\sqrt{p^j_k}\sqrt{q^j_k}e^{i(\beta^j_k-\alpha^j_k)}\right|
  \leq\sum_{k=0}^{d-1}\left|\sqrt{p^j_k}\sqrt{q^j_k}e^{i(\beta^j_k-\alpha^j_k)}\right|
  =\sum_{k=0}^{d-1}\sqrt{p^j_k}\sqrt{q^j_k}.
\end{equation}
From Eq.(\ref{DistHellinger2}) and Eq.(\ref{Bures}) we obtain
\begin{equation}\label{ineq}
    d(\Phi,\Psi)\geq D_{A^j}(\Phi,\Psi),\,\forall\Phi,\Psi\in\mathcal{H},
\end{equation}
for every $j=1,\dots,m$. Eq.(\ref{ineq}) establishes a relationship between
the Hellinger and Bures metrics. Summing the square of
Eqs.(\ref{ineq}) from $j=1$ to $j=m$ we get
\begin{equation}
    md^2(\Phi,\Psi)\geq \sum_{j=1}^m(D_{A^j}(\Phi,\Psi))^2,
\end{equation}
or, equivalently
\begin{equation}
    d(\Phi,\Psi)\geq \sqrt{\frac{1}{m}\sum_{j=1}^m(D_{A^j}(\Phi,\Psi))^2}.
\end{equation}
Therefore,
\begin{equation}
    d(\Phi,\Psi)\geq \mathcal{D}_{A^1\cdots A^m}(\Phi,\Psi),\
    \forall\Phi,\Psi\in\mathcal{H}.
\end{equation}
\qed\vspace{0.5cm}

Interestingly, Proposition \ref{distribucional} is a manifestation
of the existence of Pauli partners, because
\begin{equation}\label{implication}
   D_{A^1\cdots A^m}(\Phi,\Psi)=0\hspace{0.3cm}\mbox{does \textbf{not} imply}\hspace{0.3cm}d(\Phi,\Psi)=0,
\end{equation}
in general. In other words, two different quantum states
($d(\Phi,\Psi)\neq0$) may contain the same set of probability
distributions associated to some observables ($D_{A^1\cdots
A^m}(\Phi,\Psi)=0$). If we consider position and momentum
observables in the above explanation we get in mathematical language
the original Pauli problem \cite{Pauli}. It is important to realize
that Eq.(\ref{implication}) is not true for sets of informationally
complete observables.

\section{Quantum state reconstruction}
The physical imposition operator \cite{Goyeneche2} is a tool for
studying some geometrical problems of quantum mechanics. It has been
successfully used to find Pauli partners and maximal sets of
mutually unbiased bases \cite{Goyeneche2}, triplets of mutually
unbiased bases in dimension six \cite{Goyeneche5}, complex Hadamard
matrices in higher dimensions and SIC-POVM \cite{Goyeneche6}.
However, we could not explain before why this method is able to find
solutions to very different problems. The main objective of this
section is to present a clear explanation of the convergence of our
method from the point of view of dynamical systems theory. Our
results will be mainly restricted to the quantum state
reconstruction problem but we remark that this can be extended to
the rest of the mentioned applications straightforwardly. Let us
introduce some basic notions from dynamical systems theory
\cite{Strogatz,Arnold}.

\begin{defi}[Fixed point]
Let $\phi\in\mathcal{H}$ and $T:\mathcal{H}\rightarrow\mathcal{H}$ be a map. We say that $\phi$ is a fixed point of $T$ iff $\phi$ is invariant under $T$. That is, $T\phi=\phi$.
\end{defi}

\begin{defi}[Attractive fixed point]
Let $d(\cdot,\cdot)$ be a metric for quantum states, $T:\mathcal{H}\rightarrow\mathcal{H}$ a map and $\phi$ a fixed point of $T$. We say that $\phi$ is an attractive fixed point of $T$ if $d(T\psi,\phi)\leq d(\psi,\phi)$ for all $\psi$ contained in a neighborhood of $\phi$.
\end{defi}
One of the most famous theorems in dynamical systems is the \emph{Banach fixed point theorem}, also known as the \emph{contraction mapping theorem} or \emph{contraction mapping principle}:
\begin{teo}[Banach fixed point theorem]
Let $(\mathcal{H},d)$ be a complete metric space and $T:\mathcal{H}\rightarrow \mathcal{H}$ be a map such that
\begin{equation}\label{contraction}
d(T\Phi,T\Psi)\leq k\,d(\Phi,\Psi),
\end{equation}
for every $\Phi,\Psi\in\mathcal{H}$ and $k\in(0,1)$. Then, $T$ has a unique fixed point.
\end{teo}
This unique fixed point is always attractive and such operators are called \emph{contractions}. An operator satisfying Eq.(\ref{contraction}) for all its domain is called a \emph{Lipschitz function}, with Lipschitz constant $k$.

In what follows, we study the reconstruction of a pure state $\Phi$ from a given set of probability distributions $\{p^j_k\}$. That is, we find the complete set of pure states $\Phi$ being solution of Eq.(\ref{Born}) when the set $\{p^j_k\}$ is known. These probability distributions are associated to an informationally incomplete set of Von Neumann observables $A^1,\dots,A^m$, in general. In our study, we choose at random a state $\Phi$ (so--called the \emph{generator state}) from which we calculate the probability distributions $\{p^j_k\}$. This is in order to obtain probability distributions that can be jointly coded in a pure state. After this process, we \emph{forget} the generator state $\Phi$ and we try to reconstruct it (or one of its Pauli partners) from the knowledge of $\{p^j_k\}$. Let us formalize our definition of generator state:
\begin{defi}[Generator state]
Let $\Phi\in\mathcal{H}$ be a quantum
state and $A^1,\dots,A^m$ be a set of observables having
eigenvectors bases $\{\varphi^j_k\}$. The state $\Phi$ is called a generator state of the probability distributions $\{p^j_k\}$ if
 \begin{equation}\label{Pauli2}
    p^j_k=|\langle\varphi^j_k,\Phi\rangle|^2, k=1,\dots,N,j=1,\dots,m.
 \end{equation}
\end{defi}
Note that a generator state is not unique if the observables are
informationally incomplete. Let us define the main operator of our
work:
\begin{defi}[Physical Imposition Operator]\label{PIO}
Let $A$ be an observable having the eigenvectors basis
$\{\varphi_k\}$ and $\Phi,\Psi_0\in\mathcal{H}$. Then, we define the
Physical Imposition Operator as
\begin{equation}
    T_{A\Phi}\Psi_0=\sum_{k=0}^{d-1}
    |\langle\varphi_k,\Phi\rangle|\frac{\langle\varphi_k,\Psi_0\rangle}{|\langle\varphi_k,\Psi_0\rangle|}\varphi_k.
\end{equation}
\end{defi}
This operator is well defined for every quantum state except when $\Psi_0\perp\varphi_k$, for any $k=0,\dots,d-1$. In this case, we replace $\frac{\langle\varphi_k,\Psi_0\rangle}{|\langle\varphi_k,\Psi_0\rangle|}$
with the unity. Note that $T_{A\Phi}$ \emph{removes} the information about $A$ contained in the amplitudes of the blank state $\Psi_0$ (that is, $|\langle\varphi_k,\Psi_0\rangle|$). Additionally, $T_{A\Phi}$ \emph{imposes} the complete information about the probability distribution
\begin{equation}
p_k=|\langle\varphi_k,\Phi\rangle|^2,\hspace{0.2cm}, k=0,\dots,d-1,
\end{equation}
contained in the generator state $\Phi$. This operator is not linear and it has the following geometrical properties:
\begin{prop}\label{Prop_pio}
Let $A$ be an observable having eigenvectors basis $\{\varphi_k\}$,
$\Phi\in\mathcal{H}$ a generator state and $d(\cdot,\cdot)$ the
Bures metric. Then,
\begin{enumerate}
    \item $d(T_{A\Phi}\Psi,\Psi)\leq d(\Psi,\Phi),\,\forall\,\Phi,\Psi\in\mathcal{H}$.
    \item $d(T_{A\Phi}\Psi,\varphi_{k}) =d(\Phi,\varphi_{k}),\ \forall\,k=1,\dots,.N$.
    \item $d(T_{A\Phi}\Psi,\Phi)\leq 2\min_k d(\Phi,\varphi_{k})=2\sqrt{2}\sqrt{1-\max_{k}\sqrt{\rho_k}},\ \forall\,\Psi\in\mathcal{H}$.
    \item $d(T_{A\Phi}\Psi,\Phi)\leq2d(\Psi,\Phi),\ \forall\ \Phi,\Psi\in\mathcal{H}$.
    \item $T_{A\Phi}\xi=\xi\Leftrightarrow d(T_{A\Phi}\Psi,\xi)\leq d(\Psi,\xi),\ \forall\ \Psi\in N_{A}(\xi)$,
\end{enumerate}
where $N_{A}(\xi)$ is a neighborhood of $\xi$.
\end{prop}

 The proof of this proposition can be found in Appendix I. The Property \emph{4} relates the distance between two elements before and after applying the physical imposition operator. Interestingly, the factor 2 appearing in this inequality is a manifestation of the existence of Pauli partners. A factor less than one would mean a contradiction to existence of Pauli partners because, in this case, the physical imposition operator would be a contraction. Thus, by Banach's fixed point theorem it would have a unique fixed point. We know that this factor must be bigger than one but we do not understand why it is \emph{two} and what is the connection between this number and the maximal number of fixed points that the physical imposition operator can have (if there exists such a relationship). In order to reconstruct a quantum state we need to consider a set of observables $\{A^1,\dots,A^m\}$ having associated the set of physical imposition operators $\{T_{A^1\Phi},\dots,T_{A^m\Phi}\}$. In this case, we consider the composite physical imposition operator
\begin{equation}
T_{A^1,\dots,A^m\Phi}=T_{A^1\Phi}\circ \dots \circ T_{A^m\Phi}.
\end{equation}
The circle denoting composition will be omitted in the rest of the work. Interestingly, every state $\Psi\neq\Phi$ satisfying
\begin{equation}\label{districero}
\mathcal{D}_{A^1,\dots,A^m}(\Phi,\Psi)=0,
\end{equation}
is a Pauli partner of $\Phi$ and also a fixed point of $T_{A^1,\dots,A^m\Phi}$. This is easy to understand because $\Psi$ already contains the probability distributions that $T_{A^1,\dots,A^m\Phi}$ imposes. In particular, the generator state $\Phi$ is a fixed point of $T_{A^1,\dots,A^m\Phi}$. The complete set of states $\{\Psi\}$ satisfying Eq.(\ref{districero}) is equivalent to the complete set of solutions of the following non--linear system of coupled equations:
\begin{equation}\label{Born2}
p^j_k=|\langle\varphi^j_k,\Phi\rangle|^2.
\end{equation}
That is, the complete set of solutions of the Pauli problem. Unfortunately, the operator $T_{A^1,\dots,A^m\Phi}$ has more fixed points than Pauli partners in general. Therefore, it is very important to characterize the fixed points of the physical imposition operator.
\begin{defi}\label{Basin1}
Let $A$ be an observable and $\Phi$ a generator state. Then, the set of fixed points of  $T_{A\Phi}$ is
\begin{equation}
    \Gamma_{A\Phi}=\{\Psi\in\mathcal{H}/\, T_{A\Phi}\Psi=\Psi\}.
\end{equation}
\end{defi}
Here, we consider only one representant $\Psi$ of the ray $\Psi_{\alpha}=e^{i\alpha}\Psi$, because all of them represent the same quantum state. Now, we define a particular set of fixed points of the composite physical imposition operator.
\begin{defi}[Physical fixed points]\label{Basin2}
Let $A^1,\dots,A^m$ be a set of observables and $\Phi\in\mathcal{H}$ be a generator state. We say that
\begin{equation}
    \Gamma_{A^1\cdots A^m\Phi}=\Gamma_{A^1\Phi}\cap\cdots\cap\Gamma_{A^m\Phi},
\end{equation}
is the set of physical fixed points of $T_{A^1,\dots,A^m\Phi}$.
\end{defi}
Non--physical fixed points \emph{will not} be considered in our study
because they \emph{are not} solutions of the state reconstruction problem.
The next four propositions have a straightforward proof and their
only purpose is to clarify our recent definitions.
\begin{prop}
Let $A^1,\dots,A^m$ be a set of observables and $\Phi\in\mathcal{H}$ be a generator state. Then, the cardinality of the set of solutions of Eq.(\ref{Born2}) is given by
\begin{equation}
\mathcal{N}=|\Gamma_{A^1\cdots A^m\Phi}|.
\end{equation}
\end{prop}
As we mentioned before, the generator state $\Phi$ is a fixed point
of $T_{A^1,\dots,A^m\Phi}$. This choice lead us to
$|\Gamma_{A^1\cdots A^m\Phi}|\neq0$. In other words, the idea of
generating probability distributions from a generator state
\emph{guarantees} that $T_{A^1,\dots,A^m}$ has, at least, one
physical fixed point. Interestingly, every Pauli partner known in
literature has cardinality $\mathcal{N}<|\mathbb{N}|$ or
$\mathcal{N}=|\mathbb{R}^k|$, with $k\geq1$; but no example is known for
$\mathcal{N}=|\mathbb{N}|$, as far as we know. Roughly speaking,
Pauli partners in finite dimensional Hilbert spaces seem to form
finite or continuous sets but not infinite discrete sets.
\begin{prop}
A set of observables $A^1,\dots,A^m$ is informationally complete iff
\begin{equation}
|\Gamma_{A^1\cdots A^m\Phi}|=1,\,\forall\,\Phi\in\mathcal{H}.
\end{equation}
\end{prop}
In this case, every generator state $\Phi$ is \emph{Pauli unique}, what means that Pauli partners do not exist.
\begin{prop}
Let $A^1,\dots,A^m$ be a set of observables and $\Phi\in\mathcal{H}$ be a generator
state. Then, $\Psi\in\Gamma_{A^1\cdots A^m\Phi}$ iff $\Psi$ is a Pauli partner of $\Phi$.
\end{prop}
\begin{prop}
Let $A^1,\dots,A^m$ be a set of observables and $\Phi,\Psi\in\mathcal{H}$. Then, \begin{equation}
D_{A^1,\dots,A^m}(\Phi,\Psi)=0\,\mbox{ iff }\,\Psi\in\Gamma_{A^1\cdots A^m\Phi}.
\end{equation}
\end{prop}
Now, we present the most important result of the work:
\begin{prop}\label{attractive}
Every physical fixed point $\Psi\in\Gamma_{A^1\cdots A^m\Phi}$ is attractive for the physical imposition operator $T_{A^1\cdots A^m\Phi}$ under the Bures metric.
\end{prop}
The proof of this proposition is easy but not short, and it can be found in Appendix A. From this proposition we show that the physical imposition operator can be used to find the complete set of Pauli partners for \emph{any} set of observables and \emph{any} generator state in \emph{every} finite dimension. Proposition \ref{attractive} together with the fact that probability distributions come from a generator state are the \emph{necessary and sufficient} conditions to have convergent sequences of the form
\begin{equation}\label{sequences}
\Psi_n=(T_{A^1\cdots A^m\Phi})^n\,\Psi_0.
\end{equation}
In the case of a continuous of Pauli partners the solutions set
typically has symmetries that can be guessed by analyzing a finite set of
physical fixed points, allowing us to find the analytical expression
of the complete set of solutions. For example, we found the
complete set of Pauli partners in the case of a spin one quantum
state from considering the spin observables $S_x,S_y,S_z$, for any
generator state \cite{Weigert2}.

\begin{figure}[h!]
\label{Fig1}
\begin{center}
\subfigure[\label{Fig1a}\hspace{0.2cm} Convergence for non--complementary observables]{
\includegraphics[width=6cm]{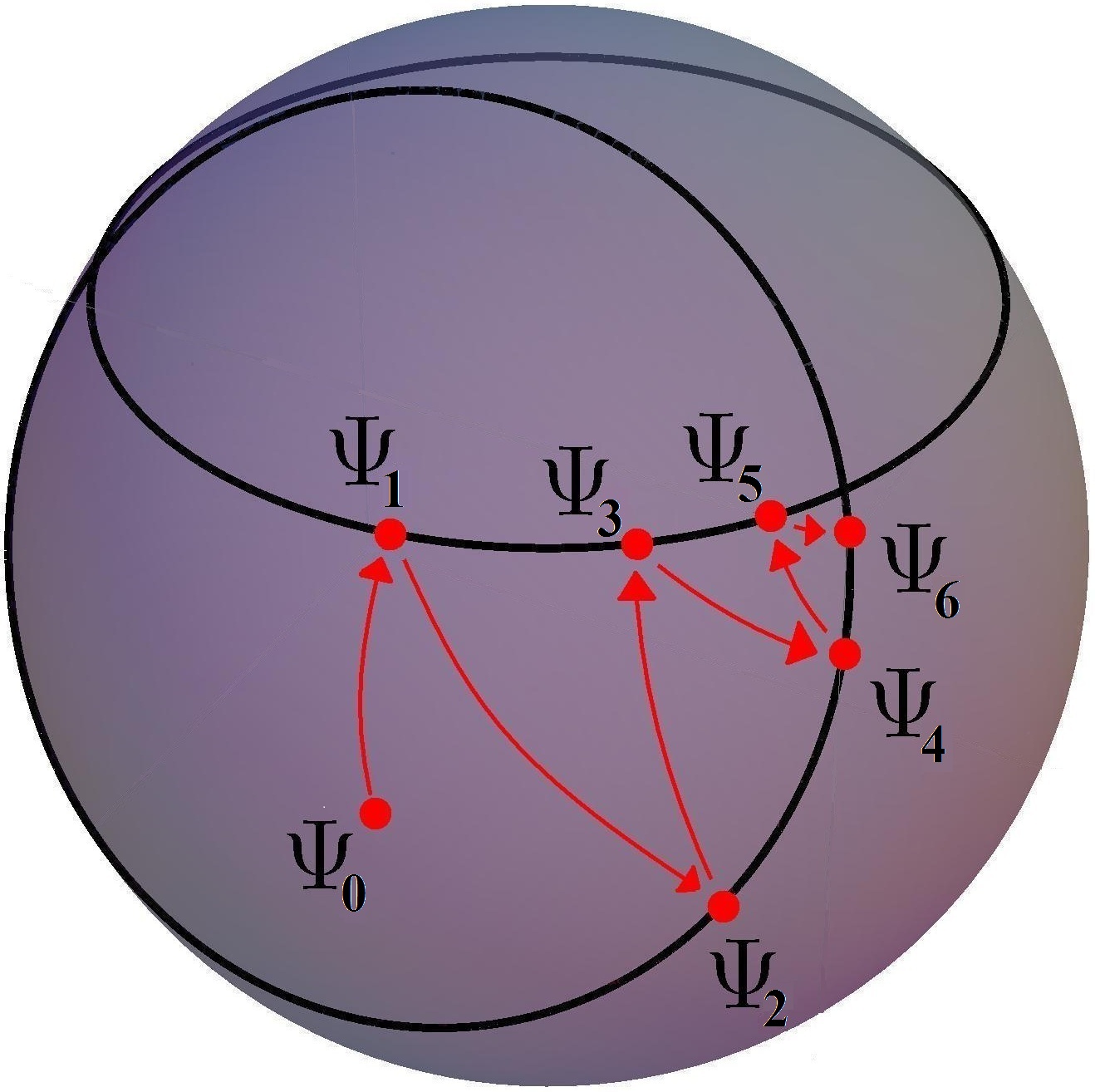}}\hspace{2cm}
\subfigure[\label{Fig1b}\hspace{0.2cm}Convergence for complementary observables]{
\includegraphics[width=6cm]{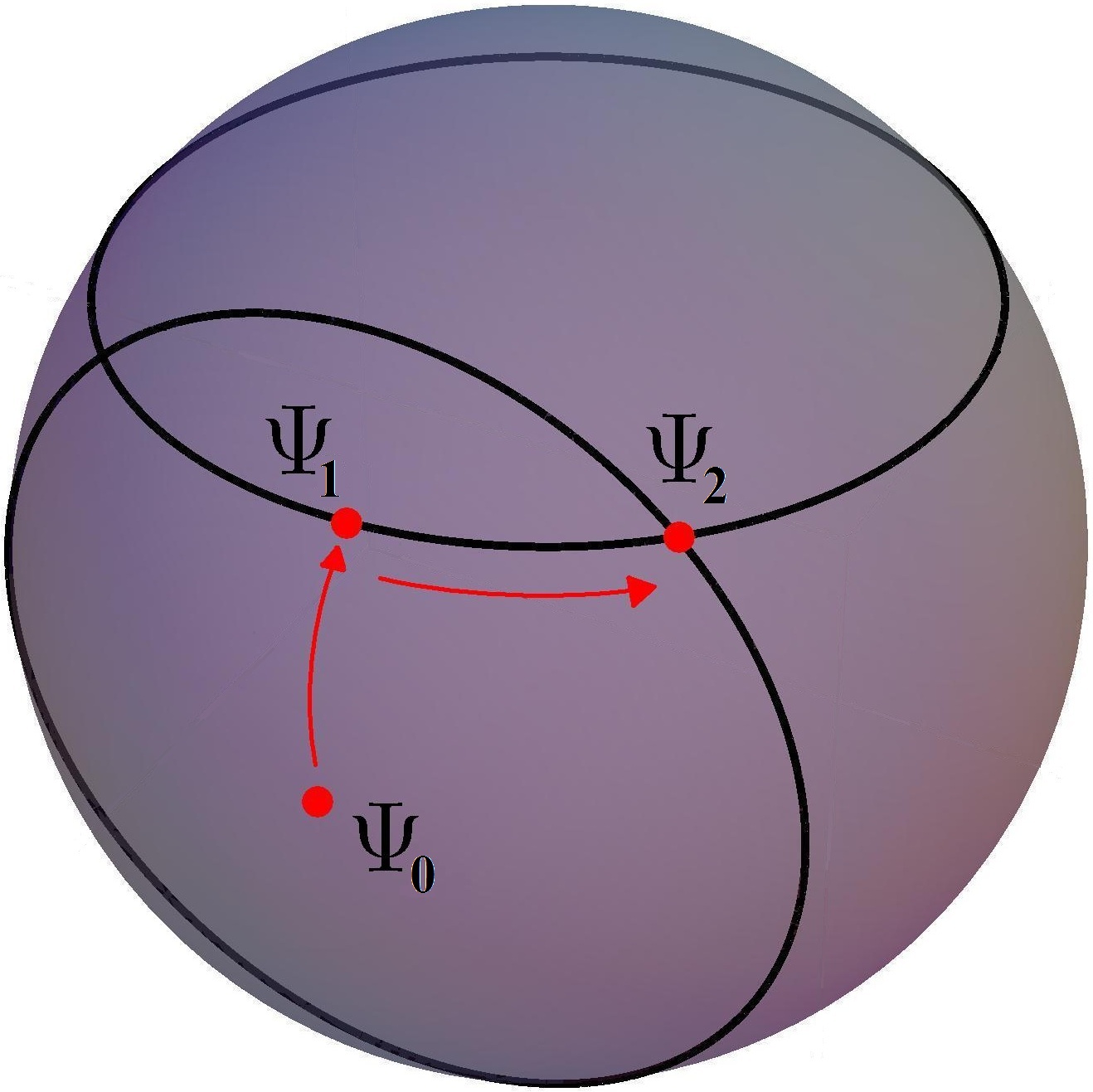}}
\caption{Bloch sphere Representation of the convergence process for the physical imposition operator.}
\end{center}
\end{figure}

In Fig.\ref{Fig1} we show the Bloch sphere representation of the
convergence process for the physical imposition operator in the case
of spin $1/2$ systems. Interestingly, the imposition operator
$T_{S_z,\Phi}$ projects the state $|\Psi_0\rangle$ onto the
horizontal circle determined by the spherical coordinates
\begin{equation}
r=\langle\Phi|\sigma_z|\Phi\rangle,\hspace{0.5cm}\sin\varphi=r,
\end{equation}
where $\varphi\in[0,\pi)$ and $\theta\in[0,2\pi)$. In Fig. \ref{Fig1a} we represent the case of two non--complementary observables, for example, $S_z$ and $S_{\eta}$, where $\eta$ is not
contained in the plane $xy$. In this figure the convergence is
clearly seen. On the other hand, if the observables are
complementary (e.g. $S_x$ and $S_z$) then the circles are embedded
in orthogonal planes and sequences converge in only two steps (see
Fig. \ref{Fig1b}). Note that every intersection of the circles is a
solution of the quantum state reconstruction problem. Thus, every
state $\Phi$ has a Pauli partner when two observables are
considered. Note that 3 observables conveniently chosen could define
3 circles such that the solution is unique. If we require uniform
probability distributions for the case of two complementary observables
then two maximal circles defined in orthogonal planes appear. Moreover, the two
intersections of these maximal circles determine a third orthogonal
basis which is mutually unbiased to the firsts two. Additionally, three maximal circles defined
in three orthogonal planes do not have an intersection point. This is consistent to the existence of a maximal set of three mutually unbiased bases in dimension two. In
higher dimensions, sequences generated by considering complementary
observables converge much faster than those generated by
non--complementary or random observables. However, convergence in a
finite number of steps is not observed. This is because we cannot define an isomorphism between the surface of the Bloch hypersphere and the set of quantum pure states when $d>2$. That is, the surface of the Bloch hypersphere has dimension $d^2-2$ and the manifold of pure states has dimension $2(d-1)$. Note that an isomorphism between these sets can be only defined for $d=2$.

A larger introduction to the physical imposition operator and a
detailed study of the convergence criteria of our method can be
found in Section III of a previous work \cite{Goyeneche5}.

\section{Bifurcations}\label{Bifurcaciones}
Let $A(\alpha)$ be a map depending on the parameters $\alpha$ and $\Gamma(A(\alpha))$ be the set of its fixed points. For some values of $\alpha$, the stability or the number of fixed points may change. In these situations, we say that the map $A$ has a bifurcation \cite{Strogatz}. In the case of the physical imposition operator the parameters are contained in the generator state $\Phi$ (considering a fixed set of observables). Thus, a bifurcation in the physical fixed points must correspond to a change in the number of physical fixed points in $\Gamma_{A^1\cdots A^m\Phi}$. This is because the stability of the physical fixed points cannot change (see Prop. \ref{attractive}). In order to clarify these ideas let us present a simple example of bifurcations for one qubit state $\Phi\in\mathbb{C}^2$: let $S_x$, $S_y$ and $S_z$ be the spin $1/2$ observables. That is,
\begin{equation}
  S_x=\frac{\hbar}{2}\left(
    \begin{array}{cc}
      0 & 1 \\
      1 & 0 \\
    \end{array}
  \right),
\hspace{0.5cm}
  S_y=\frac{\hbar}{2}\left(
    \begin{array}{cc}
      0 & -i \\
      i & 0 \\
    \end{array}
  \right),
  \hspace{0.5cm}\mbox{and}\hspace{0.5cm}
S_z=\frac{\hbar}{2}\left(
    \begin{array}{cc}
      1 & 0 \\
      0 & -1 \\
    \end{array}
  \right).
\end{equation}
Their eigenvectors bases are given by
\begin{equation}
 \mathcal{B}_x=\left\{\left(
    \begin{array}{cc}
      1 \\
      0
    \end{array}
  \right),\left(
    \begin{array}{cc}
      0\\
      1
    \end{array}
  \right)\right\},\,
  \mathcal{B}_y=\left\{\frac{1}{\sqrt{2}}\left(
    \begin{array}{cc}
      1 \\
      1
    \end{array}
  \right),\frac{1}{\sqrt{2}}\left(
    \begin{array}{cc}
      1 \\
      -1
    \end{array}
  \right)\right\},
  \,
  \mathcal{B}_z=\left\{\frac{1}{\sqrt{2}}\left(
    \begin{array}{cc}
      1 \\
      i
    \end{array}
  \right),\frac{1}{\sqrt{2}}\left(
    \begin{array}{cc}
      i \\
      1
    \end{array}
  \right)\right\},
\end{equation}
respectively. These orthonormal bases determine a maximal set of three mutually unbiased bases in $\mathbb{C}^2$. Two orthonormal bases $\{\varphi_k\}$ and $\{\phi_l\}$ of $\mathbb{C}^d$ are mutually unbiased if
\begin{equation}
|\langle\varphi_k,\phi_l\rangle|^2=\frac{1}{d},
\end{equation}
for every $k,l=0,\dots,d-1$. Let us consider the physical imposition operator $T_{S_xS_z,\Phi_1}$, where $\Phi_1\in\mathcal{B}_x\cup\mathcal{B}_z$. Then, it is easy to show that $T_{S_xS_z,\Phi_1}$ has a unique physical fixed point, up to a global unimodular factor. For example, if $\Phi_1\in\mathcal{B}_x$ then $p^x_k=\delta_{jk}$, for some $j=0,1$. Therefore, measuring the observable $S_x$ it is enough to reconstruct $\Phi$. Indeed, the state of the system is the eigenvector associated to the eigenvalue $s_j$ of the observable $S_x$. Analogously for $\Phi_1\in\mathcal{B}_z$. On the other hand, for every $\Phi_2\in\mathcal{B}_y$ we have
\begin{equation}
p^x_k=p^z_k=\frac{1}{2},
\end{equation}
for every $k=0,1$. Therefore, the operator $T_{S_xS_z,\Phi_2}$ has two physical fixed points, given by the elements of $\mathcal{B}_y$. Thus, there exists $\Phi\in\mathcal{H}$ such that $T_{S_xS_z,\Phi}$ has a bifurcation. Moreover, any continuous curve connecting $\Phi_1$ with $\Phi_2$ contains a generator state producing a bifurcation in $T_{S_xS_z,\Phi}$. Interestingly, bifurcations are manifesting that $S_x$ and $S_z$ are not informationally complete. On the other hand, if we consider the spin observables $\{S_x,S_y,S_z\}$ then the physical fixed points of $T_{S_xS_yS_z,\Phi}$ does not have bifurcations for any generator $\Phi$ (only for $d=2$). This is because maximal sets of mutually unbiased bases are informationally complete. In the case of $d=3$ the complete set of Pauli partners have been found \cite{Goyeneche2}. The example shown above can be generalized to any dimension for a general set of observables:
\begin{prop}\label{Propbifu}
A set of observables $A^1,\dots,A^m$ is informationally complete for pure states iff $T_{A^1\cdots A^m\Phi}$ has no bifurcations in its physical fixed point for any generator state $\Phi\in\mathcal{H}$.
\end{prop}
Proof: Suppose that the observables $A^1,\dots,A^m$ are informationally complete. Therefore, $T_{A^1\cdots A^m\Phi}$ has always a unique attractive fixed point for any generator state $\Phi$. Moreover, this fixed point cannot change its stability by Prop. \ref{attractive}. Then, $T_{A^1\cdots A^m\Phi}$ has no physical bifurcations. Reciprocally, suppose that $T_{A^1\cdots A^m\Phi}$ has no physical bifurcations. This means that the number of Pauli partners is the same for any generator state $\Phi$. On the other hand, it is very clear that $\Phi$ is Pauli unique when it is an eigenvector of any of the operators $A^1,\dots,A^m$. Thus, Pauli partners do not exist for any generator state $\Phi$ and $\{A^1,\dots,A^m\}$ is an informationally complete set of observables.\qed\vspace{0.5cm}
From this proposition the following corollary clearly arises: 
\begin{corol}\label{bifu}
The bifurcations of the physical imposition operators are generated in the eigenvectors of the observables.
\end{corol}
Therefore, we realize that bifurcations of the physical imposition operator are the responsible to have multiple solutions to the Pauli problem when an informationally incomplete set of observables is considered. In the case of informationally complete sets of observables there are not bifurcations. The sequence $\{\Psi_n\}$ defined in Eq.(\ref{sequences}) depends on a seed $\Psi_0\in\mathbb{C}^d$ that we choose at random in practice. This means that we need to define $2(d-1)$ random numbers and explore a sufficiently large number of seeds $\Psi_0$ in order to detect the complete set of physical fixed points. The number of seeds required depends on the generator state, the number of observables, the \emph{unbiasedness} between the eigenvectors bases of the observables and also depends on the dimension of the Hilbert space. The unbiasedness of a set of orthogonal bases is a measure of how far they are to a set of  mutually unbiased bases \cite{Durt}. Based on several numerical simulations realized in previous works we can affirm that the number of seeds $\Psi_0$ required to find the complete set of solutions:
\begin{enumerate}
  \item Increases with the dimension of $\mathcal{H}$.
  \item Decreases with the unbiasedness of the observables.
  \item Increases with the distance between the generator $\Phi$ and the closest eigenvector of the observables.
\end{enumerate}
The observables $A^1,\dots,A^m$ are parameters of $T_{A^1\cdots A^m,\Phi}$ that we usually consider as fixed. However, if we fix the generator state $\Phi$ and consider as parameters the observables $A^1,\dots,A^m$ then an interesting result arises from bifurcations theory:
\begin{prop}
If $A^1,\dots,A^m$ is an informationally incomplete set of observables then a small perturbation $A^1+\delta A^1,\dots,A^m+\delta A^m$ is also informationally incomplete.
\end{prop}
Note that the same result can be extended to informationally
complete sets of observables iff the physical imposition operator do
not have \emph{crisis} \cite{Strogatz}. It is possible to make our
method more efficient by reducing the number of free parameters of
the seed $\Psi_0$. This is stated in the following proposition:
\begin{prop}\label{seed}
Let $\Psi_0=\sum_{k=0}^{d-1}a_k\varphi_k$ be a seed, where
$\{\varphi_k\}$ is the eigenvectors basis of an observable $A^1$.
Then, the sequence
\begin{equation}
  \Psi_n=(T_{A^1\cdots A^m,\Phi})^n\Psi_0=(T_{A^m,\Phi}\circ\dots\circ T_{A^1,\Phi})^n\Psi_0,
\end{equation}
does not depend on the amplitudes of the seed $\{|a_k|\}$.
\end{prop}
This proposition has a trivial proof, because the amplitudes $\{|a_k|\}$ are missed after applying $T_{A,\Phi}$. Therefore, the number of relevant parameters of the seed $\Psi_0$ is reduced from $2(d-1)$ to $d-1$.\qed\vspace{0.5cm}

The main disadvantage of our method is that the physical imposition
operator, and all its generalizations, have more attractive fixed
points than physical ones. This means that some convergent sequences
do not correspond to a solution of our problem. Although we are able
to recognize the undesirable fixed points we cannot discard them
\emph{a priory}. This problem makes our method less efficient to find
solutions in high dimensions, where almost all the time our
numerical simulations are discarding non--physical fixed points
\emph{a posteriori}. However, we have successfully found maximal sets of mutually unbiased bases in dimension $d=31$ (992 vectors), genuine multipartite (maximally) entangled pure states up to 8 qubits and complex Hadamard matrices up to dimension 100. The problem to find a fixed point of a composite map $M=M_1\circ M_2$ which is also a fixed point of $M_1$
and $M_2$ is known as the \emph{split common fixed point problem}
\cite{Censor}. It has been proven that an iteration of convex
combinations of the form $\lambda M_1M_2+(1-\lambda)\mathbb{I}$
applied to a given seed successfully converges to a common
(attractive) fixed point of $M_1$ and $M_2$, for some kind of maps.
However, this problem is not deeply understood in the literature and it is currently an area of researching.

\section{Summary and conclusions}
In this work we studied the quantum state reconstruction problem for pure states from the point of view of dynamical systems and bifurcations theory. This novel approach allowed us to find many analytical results and also to develop a powerful tool for finding numerical solutions to geometrical problems in high dimensions. Interestingly, the rate of convergence of our method is much more faster than those obtained by standard methods. We defined the \emph{physical imposition operator} as the process of imposing information to a blank quantum state (see Def. \ref{PIO}). We demonstrated that every solution to the quantum state reconstruction problem is an attractive fixed point of this operator (see Prop. \ref{attractive}). We realized that this operator can be adapted to find solutions to many geometrical problems in Hilbert space (mutually unbiased bases, symmetric informationally complete POVM, complex Hadamard matrices, maximally entangled states, etc.). Moreover, the transition to study these problems is straightforward from the results presented in this work. On the other hand, we evidenced the existence of the quantum state reconstruction problem in four independent ways:
\begin{itemize}
\item [\emph{(i)}] \emph{Traditional approach:} Multiple solutions for a non--linear system of coupled equations does not have a unique solution (see Eq.(\ref{Born1})).
\item [\emph{(ii)}] \emph{Geometrical approach:} The inequality between the Bures and distributional metric evidences the existence of the quantum state reconstruction problem (See Prop. \ref{distribucional}).
\item [\emph{(iii)}] \emph{Dynamical systems theory approach:} The physical imposition operator is a Lipschitz operator having a Lipschitz constant 2 (see \emph{4.} in Prop. \ref{Prop_pio}). This strongly suggests the existence of the quantum state reconstruction problem.
\item [\emph{(iv)}] \emph{Bifurcations theory approach:} Bifurcations of the physical imposition operator are associated to multiple solutions to the quantum state reconstruction problem. Bifurcations only appear for informationally incomplete sets of observables (see Prop. \ref{Propbifu}). Moreover, bifurcations are generated in the eigenvectors of the observables (see Corollary \ref{bifu}).
\end{itemize}
In order to clarify ideas we explained the quantum state reconstruction problem and visualized the convergence of our method in the Bloch sphere (see Fig. \ref{Fig1}). Furthermore, the meaning of the physical imposition operator and the concept of informationally completeness were also explained from the Bloch sphere. As an example of informationally complete set we considered a maximal set of three mutually unbiased bases in dimension two. Also, we analyzed the case of two mutually unbiased bases (informationally incomplete) and interpreted the meaning of the Pauli partners in the Bloch sphere. Additionally, an explicit example of bifurcations for two--levels systems has been presented which can be also visualized in the Bloch sphere (see first part in Section \ref{Bifurcaciones}). Finally, we showed that the number of free parameters of the seeds of our method ($\Psi_0\in\mathbb{C}^d$) can be reduced from $2(d-1)$ to $d-1$ without loosing of generality in every dimension $d$ (see Prop. \ref{seed}).

\newpage
\appendix
\section{}
Here, we prove Proposition \ref{attractive}. First, we need to prove Proposition \ref{Prop_pio}. That is,
\begin{prop}
Let $A$ be an observable having eigenvectors basis $\{\varphi_k\}$,
$\Phi\in\mathcal{H}$ a generator state and $d(\cdot,\cdot)$ the
Bures metric. Then,
\begin{enumerate}
    \item $d(T_{A\Phi}\Psi,\Psi)\leq d(\Psi,\Phi),\,\forall\,\Phi,\Psi\in\mathcal{H}$.
    \item $d(T_{A\Phi}\Psi,\varphi_{k}) =d(\Phi,\varphi_{k}),\ \forall\,k=1,\dots,.N$.
    \item $d(T_{A\Phi}\Psi,\Phi)\leq 2\min_k d(\Phi,\varphi_{k})=2\sqrt{2}\sqrt{1-\max_{k}\sqrt{\rho_k}},\ \forall\,\Psi\in\mathcal{H}$.
    \item $d(T_{A\Phi}\Psi,\Phi)\leq2d(\Psi,\Phi),\ \forall\ \Phi,\Psi\in\mathcal{H}$.
    \item $T_{A\Phi}\xi=\xi\Leftrightarrow d(T_{A\Phi}\Psi,\xi)\leq d(\Psi,\xi),\ \forall\ \Psi\in N_{A}(\xi)$,
\end{enumerate}
where $N_{A}(\xi)$ is a neighborhood of $\xi$.
\end{prop}
\emph{proof:}
\begin{enumerate}
\item \begin{eqnarray}
|\langle T_{A\Phi}\Psi,\Psi\rangle|&=&\sum_{k=1}^{{N}}|\langle\varphi_{k},\Phi\rangle\langle\varphi_{k},\Psi\rangle|\\
                                    &=&\sum_{k=1}^{{N}}|\langle\langle\varphi_{k},\Psi\rangle\,\varphi_{k},\Phi\rangle|\\
                                    &\geq&\left|\left\langle\sum_{k=0}^{{N-1}}\langle\varphi_{k},\Psi\rangle\,\varphi_{k},\Phi\right\rangle\right|\\
                                    &=&|\langle\Psi,\Phi\rangle|.
\end{eqnarray}
Then,\begin{eqnarray}
    d(T_{A\Phi}\Psi,\Psi)&=&\sqrt{2}\sqrt{1-|\langle T_{A\Phi}\Psi,\Psi\rangle|}\\
                         &\leq&\sqrt{2}\sqrt{1-|\langle\Psi,\Phi\rangle|}\\
                         &=&d(\Psi,\Phi)\label{Ineq2}.
\end{eqnarray}

    \item Remembering the definition of the  physical imposition operator we can find that
\begin{eqnarray}
 |\langle T_{A\Phi}\Psi,\varphi_k\rangle|&=&\left| |\langle\varphi_{k},\Phi\rangle|\frac{\langle\varphi_{k},\Psi\rangle}{|\langle\varphi_{k},\Psi\rangle|}\right|\\
                                         &=&|\langle\Phi,\varphi_{k}\rangle|.
\end{eqnarray}
Then,
\begin{eqnarray}
    d(T_{A\Phi}\Psi,\varphi_k)&=&\sqrt{2}\sqrt{1-|\langle T_{A\Phi}\Psi,\varphi_k\rangle|}\\
                            &=&\sqrt{2}\sqrt{1-|\langle\Phi,\varphi_k\rangle|}\\
                            &=&d(\Phi,\varphi_k)\ \forall\,k=1,\dots,N.\label{eq1}
\end{eqnarray}

    \item Using the triangular inequality and Eq.(\ref{eq1})
\begin{equation}
    d(T_{A\Phi}\Psi,\Phi)\leq d(T_{A\Phi}\Psi,\varphi_k)+d(\varphi_k,\Phi)=2d(\varphi_k,\Phi), \
    \forall\,k=1,\dots,N.
\end{equation}
The most restrictive of the above inequalities is given by
\begin{equation}
    d(T_{A\Phi}\Psi,\Phi)\leq2\min_k d(\varphi_k,\Phi).
\end{equation}
The equation $$\min_k
d(\varphi_k,\Phi)=\sqrt{2}\sqrt{1-\max_k\sqrt{\rho_k}},$$ is proven immediately from the Bures metric definition.
\item From the triangular inequality and Property \emph{1.} of this proposition we have
\begin{eqnarray}
    d(T_{A\Phi}\Psi,\Phi)&\leq&d(T_{A\Phi}\Psi,\Psi)+d(\Psi,\Phi)\\
                         &\leq&2 d(\Psi,\Phi).\label{factor2}
\end{eqnarray}

\item Taking into account Eq.(\ref{Ineq2}) and considering the change of parameter
$\Psi\rightarrow\xi+\delta\xi$, $\Phi\rightarrow\xi$ we have
\begin{equation}\label{desigualdad}
    d(T_{A\Phi}(\xi+\delta\xi),\xi+\delta\xi)\leq
    d(\xi+\delta\xi,\xi).
\end{equation}
Note that in the last equation we considered
$T_{A\Phi}$ instead of $T_{A\xi}$. This is in order to cover the most general situation. That is, when $\Phi$ is a Pauli partner of $\xi$ and it is not necessarily satisfied that $\Phi=\xi$. Taking
$\xi+\delta\xi\in N_A(\xi)$ we have
\begin{eqnarray*}
  |\langle T_{A\Phi}(\xi+\delta\xi),\xi+\delta\xi\rangle|
  &=& |\langle T_{A\Phi}(\xi+\delta\xi),\xi\rangle+\langle T_{A\Phi}(\xi+\delta\xi),\delta\xi\rangle|\\
  &\approx&|\langle
  T_{A\Phi}(\xi+\delta\xi),\xi\rangle|,\label{aprox}
\end{eqnarray*}
or, equivalently
\begin{equation}
    d(T_{A\Phi}(\xi+\delta\xi),\xi+\delta\xi)\approx
    d(T_{A\Phi}(\xi+\delta\xi),\xi)\label{aprox2}.
\end{equation}
Thus, considering Eqs.(\ref{desigualdad}) and (\ref{aprox2}) we
have
\begin{eqnarray}
    d(T_{A\Phi}(\xi+\delta\xi),\xi)&\approx&
    d(T_{A\Phi}(\xi+\delta\xi),\xi+\delta\xi)\\
    &\leq&d(\xi+\delta\xi,\xi).
\end{eqnarray}
\end{enumerate}
\qed\vspace{0.5cm}

Now, we are able to give a proof of Proposition \ref{attractive}. From considering the above proposition for the observables $A^1\cdots A^m$ we have
\begin{eqnarray}
d(\Psi,\xi)&\geq& d(T_{A^1\Phi}\Psi,\xi)\\
&\geq&d(T_{A^2\Phi}T_{A^1\Phi}\Psi,\xi)\\
&\geq&d(T_{A^3\Phi}T_{A^2\Phi}T_{A^1\Phi}\Psi,\xi)\\
&\vdots&\\
&\geq&d(T_{A^m\Phi}\cdots T_{A^1\Phi}\Psi,\xi)\\
&\geq&d(T_{A^1\cdots A^m\Phi}\Psi,\xi),
\end{eqnarray}
where we consider that $\Psi\in N_{A^1\cdots
A^m}(\xi)=N_{A^1}(\xi)\cap\cdots\cap N_{A^m}(\xi).$ Then, $\xi$ is
an attractive fixed point of $T_{A^1\cdots A^m\cdots\Phi}$. Notice
that the neighborhood $N_{A^1\cdots A^m}(\xi)$ cannot be a null
measure set in state space, because each set $N_{\xi}$ contain an
open set around $\xi$. Given that an intersection of a finite number
of open sets is an open set, then $N_{ABC\cdots}(\xi)$ contain, at
least, an open set. Finally, the basin of attraction of the multiple
physical imposition operator contain the set $N_{ABC\cdots}(\xi)$.
So, its measure is not null.\qed
\end{document}